\documentclass[reprint,floatfix,superscriptaddress,showpacs,amsmath,amssymb]{revtex4-1}
\usepackage{graphicx,bm}

\usepackage{color,ulem} 
 
\usepackage{amsmath}	
\usepackage{mathrsfs}%
\begin{document} 

\title{Magnetic domain growth in a ferromagnetic Bose-Einstein 
condensate: Effects of current} 
\author{Kazue Kudo} 
\affiliation{Department of Information Sciences, 
Ochanomizu University, 2-1-1 Ohtsuka, Bunkyo-ku, Tokyo 112-8610, Japan} 
\author{Yuki Kawaguchi} 
\affiliation{Department of Applied Physics and
Quantum-Phase Electronics Center, University of Tokyo, 2-11-16 Yayoi, 
Bunkyo-ku, Tokyo 113-0032, Japan} 
 
\date{\today} 
\begin{abstract} 
 Magnetic domain patterns in a ferromagnetic Bose-Einstein condensate
 (BEC) show different properties
 depending on the quadratic Zeeman effect
 and dissipation. Another important factor that affects 
 domain patterns and
 domain growth is superfluid flow of atoms. 
 Domain growth in a ferromagnetic BEC with negative quadratic Zeeman
 energy is characterized by the same growth law as (classical)
 binary fluid in the inertial hydrodynamic regime.
 In the absence of the superfluid flow,
 the domain growth law for negative quadratic Zeeman energy is the same
 as that of scalar conserved fields such as binary alloys.    
\end{abstract} 

\maketitle 
 
\section{Introduction}
\label{sec:intro}

Domain growth laws due to domain coarsening has been studied in a wide 
variety of systems. Domain growths in ferromagnets, binary alloys, and
binary fluids are
typical examples. The equilibrium properties of a binary alloy as well as
a ferromagnet with uniaxial anisotropy can be well modeled by the Ising
model. However, the dynamics of domain coarsening in
the binary alloy, where the population
in each component is conserved, are different 
from that of non-conserved fields.  
For two-dimensional (2D) conserved scalar fields, 
the diffusive transport of the order parameter leads to the domain
growth law
$l(t)\sim t^{1/3}$, where 
$l$ is the domain size and
$t$ is time, which is
slower than the domain growth $l\sim t^{1/2}$ for non-conserved
fields~\cite{Bray,Lifshitz,Ohta,Huse}. 
The situation is more complicated in a binary fluid where fluid flow 
also
contributes to 
the transport of the order parameter.
When 
diffusion dominates domain growth,
the growth law 
reduces to
that of binary alloy: $l\sim t^{1/3}$.
However, $l(t)\sim t$ when the viscous force is dominant~\cite{Siggia}.  
If the inertia of the fluid is important, the domain size grows as 
$l(t)\sim t^{2/3}$~\cite{Furukawa}.

Magnetic domain patterns are observed in a ferromagnetic Bose-Einstein
condensate (BEC) as well as a ferromagnet.
Recent techniques for imaging of magnetization profiles in ferromagnetic 
BECs have developed to investigate the real-time  
dynamics of magnetization, such as spin texture formation 
and nucleation of spin 
vortices~\cite{Sadler2006,berkeley08,Vengalattore2010}.  
A striking feature of a ferromagnetic BEC that
is distinct from solid-state ferromagnets is that it supports
superflow of constituent atoms. The purpose of this 
paper is to find out the effect of the superflow
on the domain growth law. Spin dynamics of a ferromagnetic
BEC is well described by Gross-Pitaevskii (GP)
equation~\cite{Kawaguchi12}, which reduces to the hydrodynamic equation 
in the low energy limit.
The hydrodynamic equation has been employed
to investigate instabilities~\cite{lama,Kudo10,Yukawa},
configurations of skyrmions and spin
textures~\cite{Barnett2009,Cherng2011}, and domain pattern
dynamics~\cite{Kudo11}. 
In the presence of an energy dissipation, the hydrodynamic equation has
essentially the same form as an 
extended Landau-Lifshitz-Gilbert (LLG) equation, which describes
magnetization dynamics in conducting ferromagnets, where
the electric current in the extended LLG equation corresponds
to the superfluid current $n_{\rm tot}\bm{v}_{\rm mass}$
in the dissipative
hydrodynamic equation~\cite{Kudo11,LLG_ad,LLG_ad2}.
The analogy between the dissipative hydrodynamic equation and the
extended LLG equation has motivated us to investigate domain growth in
ferromagnetic BECs.

In this paper, we investigate the effects of $\bm{v}_{\rm mass}$ 
on the magnetic domain growth in ferromagnetic BECs for 
Ising-like cases, where longitudinal magnetization is dominant.
In our previous work, it was demonstrated that 
the superfluid flow has an effect to make
domain formation faster when the longitudinal magnetization is dominant
in a ferromagnetic BEC~\cite{Kudo11}.  
The focus of this work is on the scaling behavior of domain growth and
the growth laws. 
We use both the hydrodynamic equation and the GP equation to investigate the
magnetic domain pattern formation in (quasi-)2D systems.
By making use of the hydrodynamic equation
and artificially turning on and off the $\bm{v}_{\rm mass}$ term, we can
find out the effect of $\bm{v}_{\rm mass}$ on the domain growth
dynamics. 
We also discuss the effect of dissipation, which was originally
introduced in a phenomenological manner.  

The rest of the paper is organized as follows. 
We first introduce our model in Sec.~\ref{sec:model}.
In Sec.~\ref{sec:growth}, we briefly review
the expected domain growth laws.
Numerical simulations are given in Sec.~\ref{sec:simu}.
Our numerical simulations reproduce the expected domain growth laws.  
Conclusions are given in Sec.~\ref{sec:disc}.

\section{Model}
\label{sec:model}

We consider a spin-1 BEC of $N$ atoms under a uniform magnetic
field applied in the $z$ direction confined in a spin-independent
optical trap $U_{\rm trap}(\bm{r})$.
The mean-field energy $\mathcal{E}$ of the system is composed of the
kinetic energy $\mathcal{E}_{\rm kin}$, trapping potential energy
$\mathcal{E}_{\rm trap}$, 
the Zeeman energy $\mathcal{E}_{\rm Z}$, and the interatomic interaction
energy $\mathcal{E}_{\rm int}$: $\mathcal{E}=\mathcal{E}_{\rm
kin}+\mathcal{E}_{\rm trap}+\mathcal{E}_{\rm Z}+\mathcal{E}_{\rm int}$. 
The former three are respectively given by
\begin{align}
 \mathcal{E}_{\rm kin} &= \int d \bm{r} \sum_{m=-1}^1 \Psi_m^*(\bm{r})
\left(
-\frac{\hbar^2}{2M}\nabla^2\right) \Psi_m(\bm{r}),
\label{eq.Ekin}\\
 \mathcal{E}_{\rm trap} &= \int d \bm{r} U_{\rm trap}(\bm{r})
\sum_{m=-1}^1 |\Psi_m(\bm{r})|^2,
\label{eq.Etrap}\\
\mathcal{E}_{\rm Z} & =\int d \bm{r} \sum_{m=-1}^1 
\left(p m + q m^2 \right) |\Psi_m(\bm{r})|^2,
\label{eq.EZ}
\end{align}
where $\Psi_m(\bm{r})$ is the condensate wave function for the atoms
in the magnetic sublevel $m$, $M$ is an atomic mass,
and $p$ and $q$ are the linear and quadratic Zeeman energies per atom,
respectively. 
The wave function is normalized to satisfy 
\begin{equation}
 \int d\bm{r} \sum_{m=-1}^1 |\Psi_m(\bm{r})|^2=N.
\end{equation}
Since the total longitudinal magnetization of an atomic cloud, which is
isolated in vacuum, is conserved,  
the linear Zeeman effect merely induces the Larmor precession of atomic spins,
which can be eliminated when we move onto the rotating frame of reference.
Hence, we set $p=0$ in this paper.
On the other hand, the quadratic Zeeman energy is tunable by means of a
linearly polarized microwave field 
and takes both positive and negative values~\cite{Gerbier,Guzman2011}.

The interatomic interaction energy is given by
\begin{align}
 \mathcal{E}_{\rm int} = \frac{1}{2} \int d \bm{r} \left[c_0 n_{\rm tot}^2(\bm r) + c_1 |\bm{f}(\bm{r})|^2\right],
\label{eq.Eint}
\end{align}
where 
\begin{align}
 n_{\rm tot}(\bm{r}) &=  \sum_{m=-1}^1|\Psi_m(\bm{r})|^2,
\label{eq.n_tot} \\
f_\nu(\bm{r}) &=
 \sum_{m,n=-1}^1\Psi^*_m(\bm{r})(F_\nu)_{mn}\Psi_n(\bm{r})\ \
 (\nu=x,y,z), 
\label{eq.f_mu}
\end{align}
are the number density and the spin density, respectively, 
with $F_{x,y,z}$ being the spin-1 matrices.
The interaction coefficients $c_0$ and $c_1$ are related to the spin-$S$
{\it s}-wave scattering length $a_S$ as 
$c_0=4\pi\hbar^2(2a_2+a_0)/(3M)$ and $c_1=4\pi\hbar^2(a_2-a_0)/(3M)$.
Here, we consider only the short-range interaction and
neglect the long-range magnetic dipole-dipole interaction (MDDI),
since we are interested in the effect of the superfluid flow on the
growth dynamics of magnetic domains,  
which is dominated by the interplay between the kinetic energy and the
short-range (ferromagnetic) interaction. 
The MDDI is expected to determine the characteristic length scale of the
magnetic structure in long-time dynamics, 
and therefore, unfavorable for the study of domain growth.

We first briefly review the magnetism of spin-1 BECs.
The magnetism of spin-1 BECs is determined by the interplay between the
spin-dependent interatomic interaction [the $c_1$-term in
Eq.~\eqref{eq.Eint}] and the quadratic Zeeman effect. 
It is obvious from Eq.~\eqref{eq.Eint} that, in the absence of the
quadratic Zeeman effect, 
the condensate is fully-magnetized ($|\bm f|=n_{\rm tot}$) for a negative $c_1$,
and non-magnetized ($|\bm f|=0$) for a positive $c_1$. 
The former is referred to as the ferromagnetic phase and the latter is
the polar or antiferromagnetic phase~\cite{Ohmi1998,Ho1998}. 
Since we are interested in ferromagnetic BECs, we assume $c_1<0$ in this
paper. 
Spin-1 $^{87}$Rb atoms are known to be ferromagnetic.
The quadratic Zeeman effect introduces an easy plane ($q>0$) or an easy
axis ($q<0$) of the spontaneous magnetization. 
As seen from Eq.~\eqref{eq.EZ}, the quadratic Zeeman effect enhances the
population in the $m=0$ state for $q>0$ and those in the $m=\pm 1$ state
for $q<0$. 
Suppose that the quadratic Zeeman effect is much weaker than the
ferromagnetic interaction. 
Then, the condensate is fully-magnetized. 
Because the order parameter for a fully-magnetized state in the
direction $(\cos\alpha\sin\beta,\sin\alpha\sin\beta,\cos\beta)$ is
given by~\cite{Kudo10,Kawaguchi12} 
\begin{align}
 \bm\Psi\equiv \begin{pmatrix} \Psi_1 \\ \Psi_0 \\ \Psi_{-1}\end{pmatrix}
 = \sqrt{n_{\rm tot}}e^{i\phi} 
\begin{pmatrix}
 e^{-i\alpha}\cos^2\frac{\beta}{2} \\
 \sqrt{2}\sin\frac{\beta}{2}\cos\frac{\beta}{2} \\
e^{i\alpha}\sin^2\frac{\beta}{2}\end{pmatrix}, 
\end{align}
the population in the $m=0$ component becomes maximum at $\beta=\pi/2$,
whereas those in the $m=1$ and $-1$ component becomes maximum at $\beta=0$
and $\pi$, respectively. 
Hence, the magnetization arises in the $x$-$y$ plane for $q>0$ and in the
$+z$ or $-z$ direction for $q<0$. 
The former case corresponds to the $XY$ model and the latter the Ising
model. 

Now, we move to dynamics.
The dynamics of a spinor BEC is well described with the
time-dependent multi-component GP equation~\cite{Kawaguchi12}:
\begin{align}
 i\hbar\frac{\partial}{\partial t} \Psi_m(\bm r,t) = \frac{\delta
 (\mathcal{E} - N\mu)}{\delta \Psi^*_m(\bm r,t)}, 
\label{eq.GP1}
\end{align}
where $\mu$ is the chemical potential.
We phenomenologically introduce an energy dissipation by replacing
$i\partial /\partial t$ in Eq.~\eqref{eq.GP1} with  
$(i-\Gamma)\partial /\partial t$~\cite{Tsubota}.
Under this replacement, the chemical potential becomes time-dependent so
as to conserve the total number of atoms. 
Then, using Eqs.~\eqref{eq.Ekin}--\eqref{eq.EZ} and \eqref{eq.Eint},
we obtain the dissipative GP equation as
 \begin{align}
& (i-\Gamma)\hbar \frac{\partial}{\partial t}\Psi_m(\bm{r},t) \nonumber\\
&= 
\bigg[ -\frac{\hbar^2}{2M}\nabla^2 + U_{\rm trap}(\bm r)  -\mu(t)\nonumber\\
&\hspace{20mm} + qm^2 + c_0 n_{\rm tot}(\bm r,t)\bigg]
  \Psi_m(\bm{r},t)\nonumber\\ 
&\ \ \ \ + c_1 \sum_{n=-1}^1 \sum_{\nu=x,y,z} f_\nu(\bm
  r,t)(F_{\nu})_{mn}\Psi_n(\bm r,t). 
\label{eq.GP2}
\end{align}

We consider the low energy limit that 
the kinetic energy arising from the spatial variation of the order
parameter $\Psi_m$ is much smaller than  
both the spin-dependent ($c_1$) and spin-independent ($c_0$) interactions.
We also assume that the quadratic Zeeman energy is much smaller than the
interatomic interactions. 
Under the above assumptions, the BEC can be treated as a fully
magnetized gas, namely, the amplitude of the spin density is given by
$|\bm f|=n_{\rm tot}$ and only its direction varies in space. 
In such a situation, the physical quantities that describe the dynamics
of the ferromagnetic BEC are the normalized spin vector 
\begin{align}
 \hat{\bm f} \equiv \frac{\bm f}{n_{\rm tot}},
\end{align}
and the superfluid velocity ${\bm v}_{\rm mass}$, which is defined as
the sum of currents in all spin components: 
\begin{align}
 n_{\rm tot}\bm v_{\rm mass} \equiv \frac{\hbar}{2Mi}\sum_{m=-1}^1
 \left[\Psi_m^*(\bm\nabla\Psi_m)-(\bm\nabla\Psi_m^*)\Psi_m \right]. 
\label{eq.vmass}
\end{align}
The equations of motion for these physical quantities are derived
straightforwardly from the GP equation~\eqref{eq.GP2}. 
The detailed derivation is given in
Refs.~\cite{Kudo10,Kudo11},
and the resulting equations of motion are written as
\begin{subequations}
\begin{align}
 \frac{\partial \hat{\bm{f}}}{\partial t}
 &= \frac{1}{1+\Gamma^2}\left[
 \frac{1}{\hbar}\hat{\bm{f}} \times \bm{B}_{\rm eff}
 - (\bm{v}_{\rm mass}\cdot\nabla)\hat{\bm{f}}
 \right]
\nonumber \\
 &\quad - \frac{\Gamma}{1+\Gamma^2}\hat{\bm{f}} \times\left[
 \frac{1}{\hbar}\hat{\bm{f}} \times \bm{B}_{\rm eff}
 - (\bm{v}_{\rm mass}\cdot\nabla)\hat{\bm{f}}
 \right],
 \label{eq.extLLG}
 \\
 \bm{B}_{\rm eff} &= \frac{\hbar^2}{2M}\nabla^2\hat{\bm{f}}
 + \frac{\hbar^2}{2M}(\bm{a}\cdot\nabla)\hat{\bm{f}}
 - q\hat{f}_z\hat{z},
 \label{eq.B_eff}
 \end{align}
 \begin{align}
 M\frac{\partial}{\partial t} \bm{v}_{\rm mass} 
 &= \frac{\hbar}{2n_{\rm tot}\Gamma}\nabla\left[ 
 \nabla\cdot(n_{\rm tot}\bm{v}_{\rm mass})  \right]
\nonumber \\
&\quad
 + \hbar (\nabla\hat{\bm{f}})\cdot\left(
 \hat{\bm{f}}\times\frac{\partial\hat{\bm{f}}}{\partial t}
 \right),
 \label{eq.dvdt}
 \end{align}
\label{eq.HDE}
\end{subequations}
where $\bm{a}=(\nabla n_{\rm tot})/n_{\rm tot}$.
Here, the first term in the right-hand side of Eq.~(\ref{eq.dvdt})
does not diverge at $\Gamma=0$ since 
$\nabla\cdot(n_{\rm tot}\bm{v}_{\rm mass})\propto \Gamma$: 
Eq.~(\ref{eq.dvdt}) is also written as~\cite{Kudo11}
 \begin{align}
 M\frac{\partial}{\partial t} \bm{v}_{\rm mass} 
= - \nabla\left[ \mu_{\rm local} - \mu(t) \right]
+ \hbar (\nabla\hat{\bm{f}})\cdot\left(
 \hat{\bm{f}}\times\frac{\partial\hat{\bm{f}}}{\partial t}
 \right),
 \label{eq.dvdt.0}
 \end{align}
where
\begin{align}
 \mu_{\rm local} -\mu(t) = \frac{M}{2}\bm{v}_{\rm mass}^2
+ \frac{\hbar^2}{4M}(\nabla\hat{\bm f})^2
- \frac{\hbar^2}{2M}
\frac{\nabla^2\sqrt{n_{\rm tot}}}{\sqrt{n_{\rm tot}}}.
\end{align}

In the following sections, we investigate the spin dynamics in a
ferromagnetic BEC using both Eqs.~\eqref{eq.GP2} and \eqref{eq.HDE}. 
We discuss especially the Ising-like case ($q<0$). 
In this case, the assumption for
the spin-dependent interaction to derive the hydrodynamic equation is
weakened: we only require that the kinetic energy arising from spin
textures is negligible compared with the spin-dependent interaction 
{\it or} the quadratic Zeeman energy, since both terms favor the
ferromagnetic state. Note, however, that when the quadratic Zeeman
energy dominates the spin-dependent interaction 
($|c_1n_{\rm tot}|\lesssim |q|$)
the magnetization vanishes inside domain walls.  
We consider a quasi-2D system perpendicular to the $z$ direction (the
direction of the applied magnetic field), i.e., 
the Thomas-Fermi radius in the $z$ direction is smaller than the spin
healing length ($\xi_{\rm sp}\equiv \hbar/\sqrt{2M|c_1|n_{\rm tot}}$), 
so that magnetic structure in the $z$ direction is uniform. 
For simplicity, we neglect the confining potential in the $x$-$y$ plane
and set $\bm a=\bm 0$.

\section{Domain Growth}
\label{sec:growth}

The domain growth laws in the Ising model and binary fluids have been well
investigated experimentally, theoretically, and numerically.
Here, we briefly review the growth laws in those models especially in
two dimension, and discuss what growth laws are expected to be in
2D ferromagnetic BECs. 

The free energy for a continuum description of the  
2D Ising model is written as
\begin{align}
 F[\phi] = \int d^2x \left[ \frac12|\nabla\phi|^2 + V(\phi) \right],
\end{align}
where $\phi$ is an order parameter and $V(\phi)$ is a double-well
potential, e.g.,  
$V(\phi)=(1-\phi^2)^2$.
When the order parameter is not conserved, the time evolution 
of $\phi$ is given by the time-dependent Ginzburg-Landau 
(TDGL) equation,
\begin{align}
\frac{\partial\phi}{\partial t} = -\frac{\delta F}{\delta\phi}
=\nabla^2\phi-\frac{dV}{d\phi}.
\label{eq:TDGL}
\end{align}
This equation implies that the rate of change in the order 
parameter is proportional to the gradient of the free energy.
In this case, domain growth is driven by the surface tension 
of domain walls.  
Domain wall velocity $dl/dt$, where $l$ is the characteristic
length of domain, is approximately proportional to the curvature
$K\sim 1/l$ of the domain wall. Thus, domains grow as $l(t)\sim t^{1/2}$
at late times, when the interaction between domain walls are
negligible.  
This domain growth law for 2D non-conserved scalar fields has 
been confirmed in more sophisticated ways~\cite{Bray,Ohta}. 

When the order parameter is conserved, domain growth is 
mainly caused by the diffusive transport of $\phi$.
In this case, a continuity equation, 
$\partial\phi/\partial t=-\nabla\cdot\bm{j}$, where 
$\bm{j}=-\lambda\nabla(\delta F/\delta\phi)$, leads to
the Cahn-Hilliard equation,
\begin{align}
\frac{\partial\phi}{\partial t} 
&= \lambda\nabla^2\frac{\delta F}{\delta\phi}
\nonumber \\
&= -\lambda\nabla^2\left[\nabla^2\phi-\frac{dV}{d\phi}\right].
\label{eq:CH}
\end{align}
Here, $\lambda$ is the transport coefficient.
The Cahn-Hilliard equation is often used to describe the dynamics of
phase separation in conserved systems.
The domain growth law for 2D conserved field is $l\sim t^{1/3}$, which
has been derived in several ways~\cite{Bray,Lifshitz,Huse}.

In binary fluids, the transport of $\phi$ is caused
by hydrodynamic flow as well as diffusion. Thus, Eq.~(\ref{eq:CH})
is modified as
\begin{align}
\frac{\partial\phi}{\partial t} + \bm{v}\cdot\nabla\phi
=\lambda\nabla^2\mu,
\label{eq:CHf}
\end{align}
where $\bm{v}$ is the local fluid velocity, and 
$\mu\equiv\delta F/\delta\phi$ is the chemical potential.
The velocity obeys the Navier-Stokes equation.
In the incompressible limit, it is written by
\begin{align}
\rho\left(\frac{\partial\bm{v}}{\partial t} + 
(\bm{v}\cdot\nabla)\bm{v}\right)
= \eta\nabla^2\bm{v} - \nabla p - \phi\nabla\mu,
\label{eq:NS}
\end{align}
where $p$ is the pressure, $\eta$ is the viscosity, and the density 
$\rho$ is constant.
The left hand side of Eq.~(\ref{eq:NS}), which is inertial terms, 
vanishes in the overdamped limit.
When the inertial terms are negligible compared with the viscous force,
we shall say the system is in the viscous hydrodynamic regime.
The domain growth law is $l(t)\sim t$ in the viscous hydrodynamic 
regime~\cite{Siggia}, which has been confirmed by experiments as
well as numerical simulations~\cite{Bray}. 
In contrast, if the inertial terms are important, the system is in
the inertial hydrodynamic regime. 
The domain growth law for the inertial hydrodynamic regime is 
$l(t)\sim t^{2/3}$~\cite{Furukawa}.  
However, in classical fluids, the viscous term is usually
non-negligible, and the scaling law $l\sim t^{2/3}$ is hard to be
observed in experiments.

In a ferromagnetic BEC, if the total magnetization in the $z$-direction 
is initially zero, the $z$-component of 
magnetization is a conserved order parameter. 
Thus, if $\bm{v}_{\rm mass}=0$, the expected domain growth law 
is $l(t)\sim t^{1/3}$, which is confirmed by the 
hydrodynamic simulations for no-$\bm{v}_{\rm mass}$ cases
in Sec.~\ref{sec:simu}.
The superfluid flow in a ferromagnetic BEC corresponds to 
the hydrodynamic flow in binary liquids.
Since the viscosity vanishes in a BEC, the domain growth law in a
ferromagnetic BEC is expected to be the same as that of the inertial
hydrodynamic regime: $l(t)\sim t^{2/3}$.
This expectation is confirmed by GP and hydrodynamic simulations
is Sec.~\ref{sec:simu}.

\section{Numerical Simulations}
\label{sec:simu}

We define the characteristic length of magnetic domains by means of
scattering structure factor,
\begin{align}
 S_z(\bm{k}) = \langle\tilde{f}_z(\bm{k})\tilde{f}_z(-\bm{k})\rangle,
\end{align}
where $\langle\cdots\rangle$ represents a spacial average and 
$\tilde{f}_z(\bm{k})$
is the Fourier transform of 
$f_z(\bm{r})/n_{\rm tot}$. 
The characteristic length of magnetic domains,
i.e., domain size, $l$ is defined by $l\equiv\pi/k^*$,
where $k^*$ is the first moment of 
$S_z(k)$, 
which is the azimuth average of 
$S_z(\bm{k})$:
\begin{align}
 k^*\equiv\frac{\int_0^\infty dk\; kS_z(k)}{\int_0^\infty dk\; S_z(k)}.
\end{align}

\begin{figure}[tb]
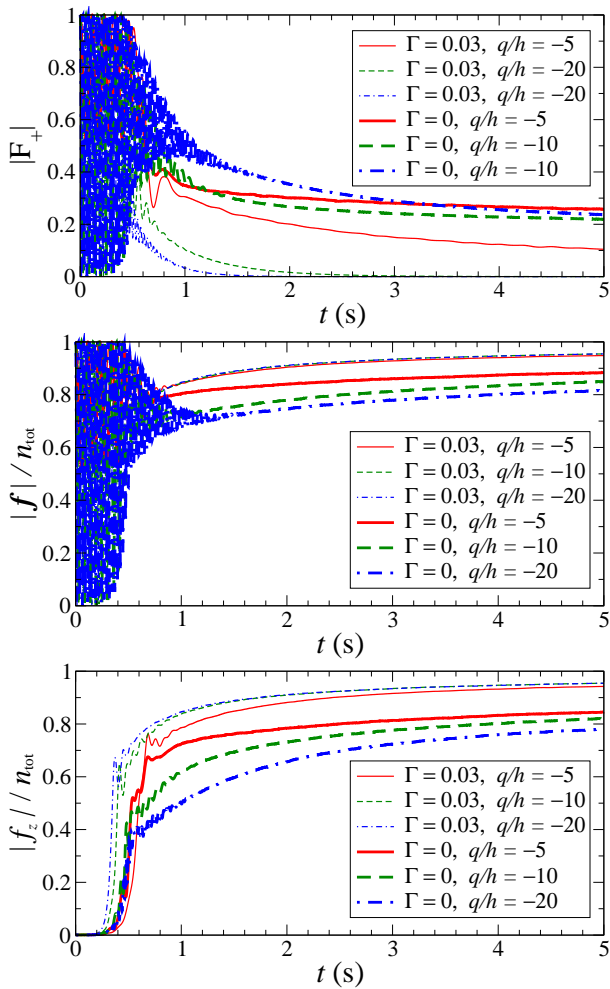

\includegraphics[width=8cm,clip]{GP_Ising_trans.eps}
\includegraphics[width=8cm,clip]{GP_Ising_mag.eps}
\includegraphics[width=8cm,clip]{GP_Ising_fz.eps}
\caption{(Color online) Time dependences of the spatial average of 
transverse magnetization
$|F_{+}|$,  
that of magnetization rate
$|\bm{f}|/n_{\rm tot}$, and that of longitudinal magnetization
$|f_z|/n_{\rm tot}$
in the GP simulations. Each curve
is the ensemble average of 5 simulations.
}  
\label{fig:GP_Ising_trans} 
\end{figure}

\begin{figure}[tb]
\includegraphics[width=8cm,clip]{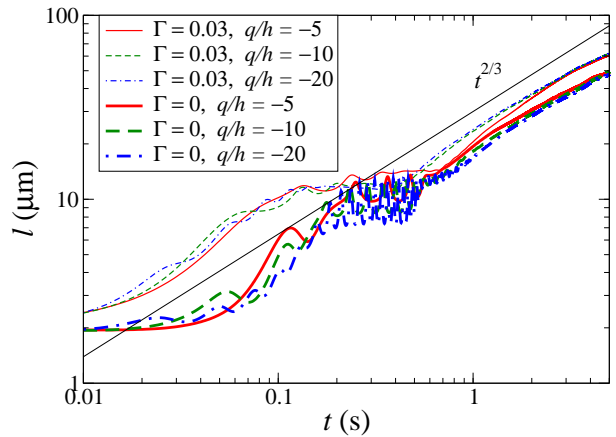}
\caption{(Color online) Time dependence of domain size $l(t)$ in
the GP simulations.  Each 
curve is the ensemble average of 5 simulations.
} 
\label{fig:GP_Ising_length} 
\end{figure}

\begin{figure}[tb]
\includegraphics[width=8cm,clip]{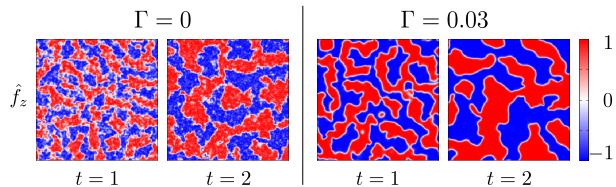}
\caption{(Color online) Snapshots of longitudinal magnetization 
$f_z/n_{\rm tot}$
obtained by the GP simulations. 
The size of each snapshot is $300$ $\mu$m on each side.
}  
\label{fig:GP_Ising_snap} 
\end{figure}

We perform numerical simulations mainly by the GP equation. Simulations by
the hydrodynamic equations are also performed to investigate the
effects of $\bm{v}_{\rm mass}$.
The mass of an atom is given by a typical value
for  a spin-1 $^{87}$Rb atom: $M = 1.44 \times 10^{-25}$ Kg.
In GP simulations, the system is in quasi-two dimension: the wave
function in the normal direction to the 2D plane is approximated by a
Gaussian with width $d$.
The total number density is taken as
$n_{\rm tot}=\sqrt{2\pi d^2}n_{\rm 3D}$ with  
$n_{\rm 3D}=2.3\times 10^{14}$ cm$^{-3}$ and $d=1$ $\mu$m.

Initial states are given as $f_x/n_{\rm tot}=1$, $f_y=f_z=0$
with additional small noises, and
periodic boundary conditions are imposed on $512$ $\mu$m $\times$ $512$
$\mu$m systems. 
The quadratic Zeeman energy is set to be a negative value so that
domains of longitudinal magnetization 
($f_z/n_{\rm tot}=\pm 1$) arise. 

In the GP simulations, we use the interaction parameters 
$c_0 n_{\rm 3D}/h = 1.3$~kHz and $c_1 n_{\rm 3D}=-5.9$~Hz.
Figure~\ref{fig:GP_Ising_trans} shows the time evolution of the spatial
average of transverse magnetization 
$|F_+|=\sqrt{f_x^2+f_y^2}/n_{\rm tot}$, 
that of magnetization rate $|\bm{f}|/n_{\rm tot}$,
and that of longitudinal magnetization $f_z/n_{\rm tot}$ obtained by the
GP simulations.  
In early times, the magnetization arises mainly in the $x$-$y$ plane and its
amplitude oscillates due to the quadratic Zeeman effect. At around
$t=0.2$--$0.6$~s, the longitudinal magnetization increases rapidly and 
the amplitude oscillations of the longitudinal magnetization appear just
after the rapid growth. The oscillations induce the oscillations of
domain size in Fig.~\ref{fig:GP_Ising_length}. 
Well-defined domains are available at $t\gtrsim 1$~s. 
The time dependence of
domain size $l(t)$ is shown in Fig.~\ref{fig:GP_Ising_length}. 
When domains are well developed at $t\gtrsim 1$~s, 
domains grow as $l(t)\sim t^{2/3}$, which 
coincide
with the domain growth law for 2D
binary fluid in the inertial hydrodynamic regime. 

The quadratic Zeeman energy determines the domain-wall structure:
the domain walls are transversely magnetized for 
$|q| \lesssim |c_1 n_{\rm 3D}|$, 
whereas the magnetization vanishes inside the domain wall for 
$|q| \gtrsim |c_1 n_{\rm 3D}|$. 
The result in Fig.~\ref{fig:GP_Ising_length} that the growth law
does not depend on the value of $q$ means that the domain-wall structure
does not affect the growth law. 
  
In Figs.~\ref{fig:GP_Ising_trans} and \ref{fig:GP_Ising_length}, we
perform calculations in both the dissipative ($\Gamma=0.03$) and
nondissipative ($\Gamma=0$) cases and obtain the same growth law 
$l(t)\sim t^{2/3}$. 
The apparent difference between those cases is the average domain size: 
the domain size of the dissipative case is larger than that of the
nondissipative case.
When domains grow, the energy associated with domain walls
decreases effectively in the dissipative case.
In other words, energy dissipation promotes domain growth.
In the nondissipative case, the decrease of the domain-wall
energy cannot be caused by dissipation, although domains can still grow
converting the domain-wall energy to other energies.
Actually, as shown in Fig.~\ref{fig:GP_Ising_trans}, the transverse
magnetization remains larger and the magnetization rate remains lower  
for the nondissipative cases than the dissipative cases, indicating that
the decrease in the domain-wall energy is compensated with the increase
in the quadratic Zeeman energy and the ferromagnetic interaction
energy.

Snapshots of longitudinal magnetization 
$f_z/n_{\rm tot}$ in both dissipative
($\Gamma=0.03$) and nondissipative ($\Gamma=0$) cases are demonstrated
in Fig.~\ref{fig:GP_Ising_snap}.  
The snapshot for $\Gamma=0$ at $t=2$, where small white fractions
still remains inside domains, reflects the fact that
transverse magnetization does not vanish even in late times for 
$\Gamma=0$.

\begin{figure}[tb]
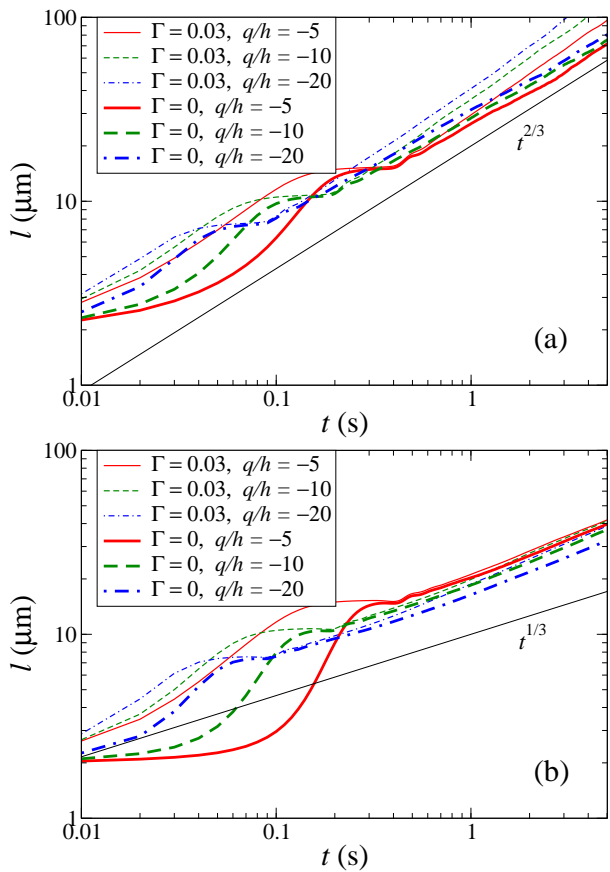

\includegraphics[width=8cm,clip]{HD_Ising_length.eps}
\includegraphics[width=8cm,clip]{HD_Ising_length_v0.eps}
\caption{(Color online) Time dependence of domain size $l(t)$ 
obtained by the hydrodynamic simulations 
(a) with $\bm{v}_{\rm mass}$ and (b) without $\bm{v}_{\rm mass}$.  
Each curve
is the ensemble average of 5 simulations.
}
\label{fig:HD_Ising_length} 
\end{figure}

\begin{figure}[tb]
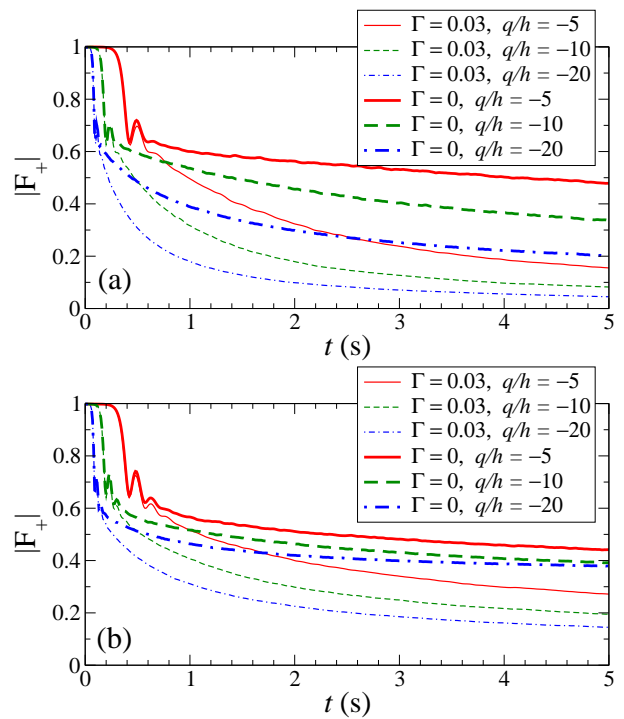

\includegraphics[width=8cm,clip]{HD_Ising_trans.eps}
\includegraphics[width=8cm,clip]{HD_Ising_trans_v0.eps}
\caption{(Color online) Time dependences of the spatial average of 
transverse magnetization $|F_{+}|$ in (a) fully-hydrodynamic 
 and (b) no-$\bm{v}_{\rm mass}$ simulations. 
Each curve
is the ensemble average of 5 simulations.
} 
\label{fig:HD_Ising_trans} 
\end{figure}

\begin{figure}[tb]
\includegraphics[width=8cm,clip]{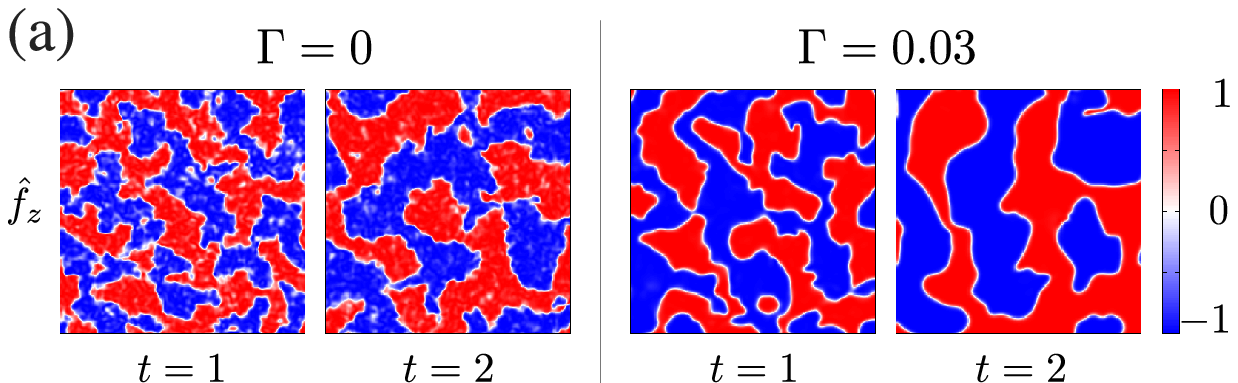}
\includegraphics[width=8cm,clip]{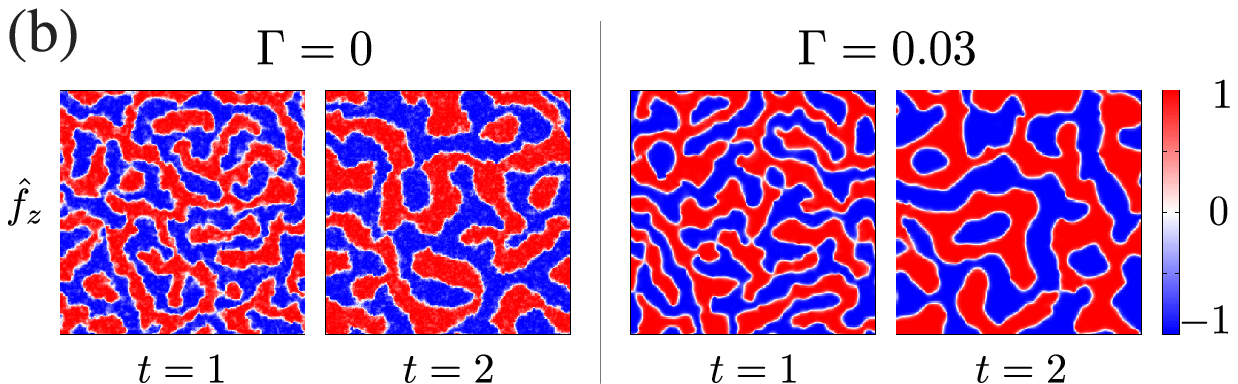}
\caption{(Color online) Snapshots of longitudinal magnetization
 $\hat{f}_z$ in (a) fully-hydrodynamic 
 and (b) no-$\bm{v}_{\rm mass}$ simulations.
The size of each snapshot is $300$ $\mu$m on each side.
} 
\label{fig:HD_Ising_snap} 
\end{figure}

In order to investigate the effects of superfluid flow, we have
performed simulations by means of the 
hydrodynamic equations in the presence
and in the absence of $\bm{v}_{\rm mass}$. 
The time dependences of domain size $l(t)$ in fully-hydrodynamic
and no-$\bm{v}_{\rm mass}$ simulations are shown in
Figs.~\ref{fig:HD_Ising_length}(a) and \ref{fig:HD_Ising_length}(b),
respectively.  
In the fully-hydrodynamic cases, as seen in
Fig.~\ref{fig:HD_Ising_length}(a), domains grow as $l(t)\sim t^{2/3}$,
which is the same growth law as in the GP simulations. 
Figure~\ref{fig:HD_Ising_length}(b) indicates that the growth law in
no-$\bm{v}_{\rm mass}$ cases coincides with that of 2D conserved 
scalar fields: $l(t)\sim t^{1/3}$. 
The dependences on $q$ and $\Gamma$ are not noticeable.

The time dependence of transverse magnetization in the hydrodynamic
simulations, which is shown in Fig.~\ref{fig:HD_Ising_trans}, shows
similar tendency to that in the GP simulations: the transverse
magnetization remains larger in the nondissipative cases than the
dissipative cases. 
The difference between dissipative and nondissipative cases is larger
in Fig.~\ref{fig:HD_Ising_trans}(a) than
Fig.~\ref{fig:HD_Ising_trans}(b). 
Since the magnitude of transverse magnetization is related with the
loss of the domain-wall energy,
this is consistent with the results
in Fig.~\ref{fig:HD_Ising_length}, i.e., 
the difference in domain size between dissipative and nondissipative
cases is larger in the fully-hydrodynamic simulations than the 
no-$\bm{v}_{\rm mass}$ simulations.

Figure~\ref{fig:HD_Ising_snap} shows the snapshots of $\hat{f}_z$ in
(a) fully-hydrodynamic 
and (b) no-$\bm{v}_{\rm mass}$ simulations.
Domain size is apparently larger in (a) than (b), which is in good
agreement with
the results in Fig.~\ref{fig:HD_Ising_length}. 

Finally, we discuss the effect of a confining potential, which is
unavoidable in experiments, on the growth law. 
The effect of the confining potential is to cause a non-uniform
density profile $n_{\rm tot}$. Since the spin-independent interaction is
much stronger than the spin-dependent interaction, this density
distribution is not affected by spin structures, and thus, does not
evolve in time. Such a time-independent term does not contribute to the
scaling behavior of domain growth. Hence, the growth law is expected not
to change even in the presence of a confining potential. 

\section{Conclusions}
\label{sec:disc}

We have performed the GP and hydrodynamic simulations to investigate
the effects of the superfluid flow on the magnetic domain growth in
ferromagnetic BECs with negative quadratic Zeeman energies. 
Domain patterns formed by the longitudinal magneization grow as time
evolves.
The characteristic domain size grows, as expected, as
$l(t)\sim t^{2/3}$ in the GP and the fully-hydrodynamic simulations and 
$l(t)\sim t^{1/3}$ in the hydrodynamics simulations without
the superfluid flow. 
The former and the latter correspond to the growth
laws for diffusive and inertial hydrodynamic dynamics in conserved
scalar fields, respectively.
The domain growth laws are almost independent of the quadratic
Zeeman energy and dissipation, although domain size increases
earlier in the dissipative 
case than the nondissipative case especially in the GP and the
fully-hydrodynamic simulations.

In conclusion, the superfluid flow promotes the magnetic domain growth
in ferromagnetic BECs. The growth laws 
for negative quadratic Zeeman energy
are the same as those of conserved scalar fields 
and are independent of the quadratic Zeeman energy and 
dissipation.

\begin{acknowledgments}
This work was supported by KAKENHI (22340114, 22740265, 22103005) from
 MEXT of Japan, and by Funding  
Program for World-Leading Innovation R \& D on Science
and Technology (FIRST). YK acknowledges the financial support from Inoue
 Foundation for Science. 
\end{acknowledgments}

\end{document}